\documentclass[a4paper,twocolumn,english,8pt,amssymb,nofootinbib,superscriptaddress]{revtex4}
\usepackage{lmodern}
\usepackage{lmodern}

\usepackage[T1]{fontenc}
\usepackage[latin9]{inputenc}
\usepackage{babel}
\usepackage{mathrsfs}
\usepackage{amsmath}
\usepackage{amssymb}
\usepackage{esint}
\usepackage[unicode=true,pdfusetitle,
bookmarks=true,bookmarksnumbered=false,bookmarksopen=false,
breaklinks=false,pdfborder={0 0 1},backref=false,colorlinks=false]
{hyperref}

\makeatletter

\@ifundefined{textcolor}{}
{%
	\definecolor{BLACK}{gray}{0}
	\definecolor{WHITE}{gray}{1}
	\definecolor{RED}{rgb}{1,0,0}
	\definecolor{GREEN}{rgb}{0,1,0}
	\definecolor{BLUE}{rgb}{0,0,1}
	\definecolor{CYAN}{cmyk}{1,0,0,0}
	\definecolor{MAGENTA}{cmyk}{0,1,0,0}
	\definecolor{YELLOW}{cmyk}{0,0,1,0}
}

\usepackage{babel}
\usepackage{babel}
\usepackage{babel}
\usepackage{babel}
\usepackage{babel}
\usepackage{babel}
\usepackage{babel}
\usepackage{babel}
\usepackage{babel}
\usepackage{babel}
\usepackage{babel}
\usepackage{babel}
\usepackage{babel}
\usepackage{babel}
\usepackage{babel}
\usepackage{babel}

\usepackage{babel}
\usepackage{graphicx}
\def\b{\begin{equation}}
\def\e{\end{equation}}

\@ifundefined{textcolor}{}{%
	\definecolor{BLACK}{gray}{0}
	\definecolor{WHITE}{gray}{1}
	\definecolor{RED}{rgb}{1,0,0}
	\definecolor{GREEN}{rgb}{0,1,0}
	\definecolor{BLUE}{rgb}{0,0,1}
	\definecolor{CYAN}{cmyk}{1,0,0,0}
	\definecolor{MAGENTA}{cmyk}{0,1,0,0}
	\definecolor{YELLOW}{cmyk}{0,0,1,0}
}

\usepackage{latexsym}\usepackage{bm}

\makeatother

\begin{document}
	\title{Spherical timelike orbits around Kerr black holes }
	
	\author{Aydin Tavlayan}
	
	\email{aydint@metu.edu.tr}
	
	\selectlanguage{english}%
	
	\affiliation{Department of Physics,\\
		Middle East Technical University, 06800 Ankara, Turkey}
	\author{Bayram Tekin}
	
	\email{btekin@metu.edu.tr}
	\affiliation{Department of Physics,\\
		Middle East Technical University, 06800 Ankara, Turkey}

	\selectlanguage{english}%
\begin{abstract}
\noindent We study the order ten polynomial equation satisfied by the radius of the spherical timelike orbits for a massive particle with a generic energy around a Kerr black hole. Even though the radii of the prograde and retrograde orbits at the equatorial or polar plane for particles with zero or unit energy have a monotonic dependence on the rotation parameter of the black hole, we show that there is a critical inclination angle above which the retrograde orbits have a nonmonotonic dependency on the rotation of the black hole. Thus the capture cross-section of these retrograde orbits decrease with increasing black hole spin. Hence their efficiency to reduce the black hole's spin is decreased.  We also provide an equation for the critical inclination angle that shows exactly at which point the nonmonotonicity starts. In addition, we employ the Lagrange-B\"{u}rmann method to find approximate analytic solutions from the known exact solutions.
\end{abstract}
\maketitle

\section{Introduction}

\noindent Recently in \cite{Aydin}, we studied the spherical light orbits (that is constant radius orbits)  around Kerr black holes \cite{Kerr}. Radii of the two equatorial orbits (one prograde and one retrograde) and the polar orbit (that is  the zero angular momentum orbit which is dragged by the black hole) had been known analytically for some time. All these three orbits are outside the event horizon and they are unstable. There is also another polar orbit inside the event horizon given by equation (17) in \cite{Aydin} which is a saddle point. [For completeness, we give these solutions in Appendix A]. To extend this family of known four analytic solutions for light orbits, we found solutions in between these equatorial and polar extremes. One particularly interesting finding was that as opposed to the equatorial orbits, the radius of the photon sphere for these new spherical orbits does not depend monotonically in the rotation of the black hole. This result might have implications on the maximum value of the spin of a black hole which we shall discuss a little bit below. 

\noindent In the same work, we also introduced the powerful Lagrange-B\"{u}rmann technique to find very accurate numerical solutions of generic spherical orbits using the known analytical solutions.  In this current work, we shall carry out a similar analysis for the case of massive particles following timelike spherical orbits. The problem gets much more complicated for two different reasons: first the relevant polynomial determining the constant radius light orbits is a {\it sextic} while for massive particles it is an {\it order ten} polynomial; and secondly the constant radius light orbits do not depend on the energy of the light, whereas energy is relevant for massive particles.  So the parameter space in the case of the massive spherical orbits is larger and the ensuing discussion becomes harder. 

\noindent Motion of light and  massive particles around a black hole reveal a great deal of information about the black hole. The superb image of the supermassive black hole at the center of the M87 galaxy, provided by the Event Horizon Telescope, is a testament to this \cite{EHT}. A particle moving in the environment of a rotating Kerr black hole, besides its mass ($\mu$), has three conserved quantities: its energy $E$,  one component of its angular momentum $L_z$, and the Carter's constant ${\mathcal{Q}}$ which represents motion perpendicular to the equatorial plane \cite{Carter}. The first two follow from spacetime Killing vectors, while the last one follows from a symmetric rank-2 tensor. Among the orbits around a rotating black hole, constant radius orbits  are particularly useful to understand \cite{Teo1, Cunha,Teo2}.
 
\noindent Let us briefly recapitulate the main points in the case of null geodesics \cite{Aydin} as they will be relevant to the current work.
Let the mass of the black hole be $m$ and its rotation parameter be $a := J/m$ with $J$ being the angular momentum of the black hole. Assuming the  bound light/photon moving in a spherical orbit has a conserved $L_z$ angular momentum, then one defines the parameter $\nu$ as (which is related to the effective inclination angle as will be seen in the next section) 
\begin{equation}
\nu := \frac{ \mathcal{Q}}{\mathcal{Q}+ L_z^2}.
\end{equation}
Then the radius $r$ of an arbitrary photon sphere satisfies a {\it sextic } polynomial in such a way that one has  $r= m f( u, \nu)$ with $u:= \frac{ a^2}{m^2}$. So, the energy of the photons are 
not relevant due to the equivalence principle since photons with different energies can move in the same orbit. As mentioned above and as given in Appendix A, the function $f$ had been known for $\nu =0$ (one prograde and  one retrograde equatorial orbit outside the black hole) and for $\nu =1$ (one polar orbit outside the black hole) and in \cite{Aydin} we found a critical $\nu_{\text{cr}}$ for which the prograde and retrograde solutions can be constructed analytically as the sextic factors to a quadratic times a quartic. This critical $\nu_{\text{cr}}$ is a function of the rotation parameter $u$ given as 
\begin{equation}\label{criticalphoton}
\nu_{\text{cr}}= \frac { 3 \left [ 1- (1-u)^{\frac{1}{3}} \right]^3}{ 7 u + u (1-u)^{\frac{1}{3}}\left ( (1-u)^{\frac{1}{3}}-5\right )},
\end{equation}
such that  $\nu_{\text{cr}} \in [0, \frac{3}{7}]$. So for each rotating black hole, there are generically six more analytical solutions besides the known four solutions for photon orbits. The equatorial prograde and retrograde orbits depend {\it monotonically } on the rotation parameter of the black hole; but the critical retrograde orbit is nonmonotonic. This surprising result was first observed for one particular   $\nu_{\text{cr}}= 3/7$ in \cite{Hod} and was extended to the domain  $\nu_{\text{cr}} \in [0, \frac{3}{7}]$ in \cite{Aydin}. 

\noindent This nonmonotonic behavior of the non-equatorial retrograde orbits on the rotation parameter is important for the following reason. Assume that a black hole is rotating very closely to the extremal limit, namely $J$ is slightly lower than $m^2$. One might envisage a devilish experiment and throw the Earth (with $J_E/M_E^2 \approx 90$) into that black hole to overspin it  and make it lose its event horizon to reveal its singularity. The Cosmic Censorship conjecture would be violated and this would be disastrous. But as Thorne \cite{Thorne} showed, it is not possible to overspin a black hole using such a scheme simply because the highly spinning object (in our experiment the Earth) would disintegrate all the way to a plasma and start radiating photons in all directions. Retrograde photons in the equator of the black hole has a larger impact parameter than the prograde photons and hence they are captured more, leading to a slow down of the spin of the black hole. In fact, making some assumptions, Thorne found that the {\it maximum} spin of the black with such an equatorial accretion scheme would be $J \approx 0.998 m^2$ which is smaller than the extremal case. But as mentioned, the impact parameter for the equatorial orbits are monotonically increasing function of the rotation of the black hole, while the generic spherical orbits are nonmonotonic, hence, for non-equatorial accretion processes one must re-compute Thorne's maximum spin. For this purpose our results in \cite{Aydin} and here will be relevant.  

\noindent In the case of timelike orbits, radius of the orbit depends on $three$ parameters instead of two: one has  $ r = m f ( u, \nu, k)$ where  $u$ and $\nu$ are defined as above and $k$ is a dimensionless quantity related to the energy of the massive particle which we shall describe below. 

\noindent The layout of this work is as follows: In section 2, we obtain the generic order ten polynomial for the spherical timelike orbits around the Kerr black hole. In section 3, we discuss the Schwarzschild limit. In section 4, we study the equatorial orbits for the Kerr black hole. In section 5, we study the zero energy solutions. In section 6, we study the nonmonotonic behavior of the orbits as a function of the black hole rotation parameter for nonzero energy. In section 7, we study the unit energy solutions. In Appendix A, we revisit the null orbit case, in Appendix B we described the Lagrange-B\"{u}rmann theorem. In Appendix C, we define the discriminant of a higher order polynomial when the roots are not known. In Appendix D, we study bound and unbound orbits with generic energy values.

\section{Timelike Geodesics}

\noindent In the standard Boyer-Lindquist coordinates $(t,r,\theta,\phi)$, the Kerr black hole metric (in the $G=c=1$ units) can be written as
\begin{eqnarray}
ds^2&=&-\left(1-\frac{2mr}{\Sigma}\right)dt^2-\frac{4mr}{\Sigma}a\sin^2{\theta} dt d\phi  \nonumber\\
&&+\Sigma\left(\frac{dr^2}{\Delta}+d\theta^2\right)\\
&&+\left(r^2+a^2+\frac{2mr}{\Sigma}a^2 \sin^2{\theta}\right)\sin^2{\theta}d\phi^2, \nonumber
\end{eqnarray}
where 
\begin{eqnarray}
&&\Sigma:=r^2+a^2 \cos^2{\theta}, \\
&&\Delta:=r^2-2mr+a^2. \nonumber
\end{eqnarray}
The relevant geodesic equation governing the radial motion of a particle with mass $\mu$ is
\begin{eqnarray}
&&\Sigma\frac{dr}{d\tau}=\pm \sqrt{R(r)},
\end{eqnarray}
with 
\begin{eqnarray}
R(r):= &&r^2 \left[a^2 \left(E^2-\mu ^2\right)-{\mathcal{Q}}-L_z ^2\right]\\
&&-a^2 {\mathcal{Q}}+2 m r \left[(a E-L_z )^2+{\mathcal{Q}}\right]\nonumber \\
&&+r^4 \left(E^2-\mu ^2\right)+2 \mu ^2 m r^3. \nonumber
\end{eqnarray}
Here, $\tau$ is an affine parameter along the geodesic. $E$, $L_z$ and ${\mathcal{Q}}$ are constants of motion mentioned in the Introduction.  Since we already studied the $\mu=0$ case in \cite{Aydin}, we shall consider $\mu \ne 0$ which then suggests the following rescalings:
\begin{eqnarray}
&&r = m x, \hskip 0.5 cm a = m \tilde{a}, \hskip 0.5 cm {\mathcal{Q}} = m^2 \mu^2 \tilde{{\mathcal{Q}}}, \\ 
&&L_z = m \mu \tilde{L}_z, \hskip 0.5cm E = \mu \tilde{E}, \nonumber
\end{eqnarray}
which remove the mass $(m, \mu)$ dependence of the radial geodesic equation and allow us to work with dimensionless parameter, as expected: 
\begin{eqnarray}
\frac{R(r)}{\mu^2 m^4}= &&x^2 \left[\tilde{a}^2 \left(\tilde{E}^2-1\right)-\tilde{{\mathcal{Q}}}-\tilde{L}_z ^2\right]-\tilde{a}^2 \tilde{{\mathcal{Q}}}\\
&&+2  x \left[(\tilde{a} \tilde{E}-\tilde{L}_z )^2+\tilde{{\mathcal{Q}}}\right] \nonumber \\
&&+x^4 \left(\tilde{E}^2-1\right)+2 x^3. \nonumber
\end{eqnarray}
By defining the conserved {\it effective inclination angle} as \cite{Ryan}
\begin{equation}
\cos i := \frac{L_z}{\sqrt{ L_z^2 + {\cal Q}}}, \label{inclination}
\end{equation}
and
\begin{equation}
\nu:=\sin^2{i},
\end{equation}
the Carter's constant can be put into the form
\begin{equation}
 {\cal Q}=\frac{\nu L_{z}^{2}}{1-\nu}. \label{Relation1}
\end{equation}
Using this in the following conditions for spherical geodesics
\begin{equation}
R(r) =0, \hskip  0.5 cm \frac{ d R}{dr}=0,
\end{equation}
and with the help of the redefinitions
\begin{equation}
u=\tilde{a}^2, \hskip 0.5 cm k=\tilde{E}^2,
\end{equation}
the Carter's constant and the $z$-component of the angular momentum of the particle can be eliminated and the following polynomial equation can be obtained
\begin{eqnarray}\label{polynomial}
p(x,u,\nu,k)&=&x^{10}\left(k-1\right)^2\nonumber \\
&+&2x^9(k-1) (4-3 k)\nonumber\\
&+&x^8 \left[4 \left(k-1\right)^2 \nu  u+9 k^2-32 k+24\right]\nonumber\\
&+&x^7 \left[-2 \left(k-1\right) u \left(k (6 \nu +2)-11 \nu -1\right)\right.\nonumber\\
&&\hskip 0.8 cm \left.+24 k-32\right]\nonumber\\
&+&x^6 \left[4 k^2 \nu ^2 u^2-8 k \nu ^2 u^2+2 k^2 \nu  u^2-4 k \nu  u^2\right. \nonumber\\
&&\hskip 0.8 cm \left. +2 u \left(6 k^2 \nu -k (23 \nu +5)+20 \nu +4\right)\right.\nonumber\\
&&\hskip 0.8 cm \left.+4 \nu ^2 u^2+2 \nu  u^2+16\right]\nonumber\\
&+&x^5 \left[12 k \nu ^2 u^2-12 k^2 \nu  u^2+24 k \nu  u^2+16 k \nu u\right.\nonumber\\
&&\hskip 0.8 cm \left.-12 \nu ^2 u^2-12 \nu  u^2\right.\nonumber\\
&& \hskip 0.8 cm \left. -24 \nu  u-8 u\right]\nonumber\\
&+&x^4 \left[4 k^2 \nu ^2 u^3-12 k^2 \nu ^2 u^2-8 k \nu ^2 u^3\right.\nonumber\\
&&\hskip 0.8 cm \left. +12 k \nu ^2 u^2+10 k^2 \nu  u^2\right.\nonumber\\
&&\left.-28 k \nu  u^2+4 \nu ^2 u^3+9 \nu ^2 u^2+14 \nu  u^2+u^2\right]\nonumber\\
&+&x^3 \left[4 k^2 \nu ^2 u^3+2 k \nu ^2 u^3-16 k \nu ^2 u^2\right.\nonumber\\
&&\hskip 0.8 cm \left.-4 k^2 \nu  u^3+6 k \nu  u^3 \right.\nonumber\\
&& \hskip 0.8 cm \left.+8 k \nu  u^2-6 \nu ^2 u^3-2 \nu  u^3\right]\nonumber\\
&+&x^2 \left[k^2 \nu ^2 u^4-4 k^2 \nu ^2 u^3-2 k \nu ^2 u^4\right.\nonumber\\
&&\hskip 0.8 cm \left.+10 k \nu ^2 u^3-2 k \nu  u^3+\nu ^2 u^4\right]\nonumber\\
&+&2 x(k-1) k \nu ^2 u^4\nonumber\\
&+&k^2 \nu ^2 u^4=0. 
\end{eqnarray}
This is the main equation that we shall study; generically it has 10 roots and since the event horizon is located at $x_\text{H} = 1+ \sqrt{1-u}$, ideally, one would like to find all viable real roots, $x > x_\text{H}$, of the form $x=x(u, \nu, k)$. But, this polynomial equation cannot be solved analytically in terms of a  finite number of radicals as stated in the Abel-Ruffini impossibility theorem. Therefore, below, some specific cases are investigated. Let us first note that, as $k$ is basically the square of the energy of the massive particle as measured from a far-away observer, $ 0\le k <1$ solutions are bound solutions, while $k=1$ yields marginally bound solutions and $k >1$ gives unbound solutions.  

\noindent Let us note that in the limit $ k\rightarrow \infty$,  (\ref{polynomial}) reduces to the one obtained for null geodesics which is equation (9) in \cite{Aydin} 
\begin{eqnarray}
p(x,u,\nu, k \rightarrow \infty)&= x^6-6 x^5+(9+2 \nu u) x^4-4 u x^3 \nonumber \\
&-\nu  u (6 -u) x^2+2 \nu  u^2 x +\nu  u^2=0,\nonumber\label{sextic0}
\end{eqnarray} 
which we studied in the mentioned reference in great detail.

\section{The Schwarzschild Limit}
\noindent The simplest case to start with is the Schwarzschild limit where the rotation parameter is set zero, $u=0$. The polynomial (\ref{polynomial}) simplifies to
\begin{equation}
p(x,u=0,\nu, k )=x^6 \left(k x^2-3 k x-x^2+4 x-4\right)^2,
\end{equation}
with nontrivial roots 
\begin{eqnarray}
&& x_1(k)=\frac{-3 k+ \sqrt{9 k^2-8k}+4}{2-2 k}, \nonumber \\
&&x_2(k)=\frac{8}{-3 k+ \sqrt{9 k^2-8k}+4}. 	
\label{Schw}
\end{eqnarray}

\begin{figure}
	\centering
	\includegraphics[width=1\linewidth]{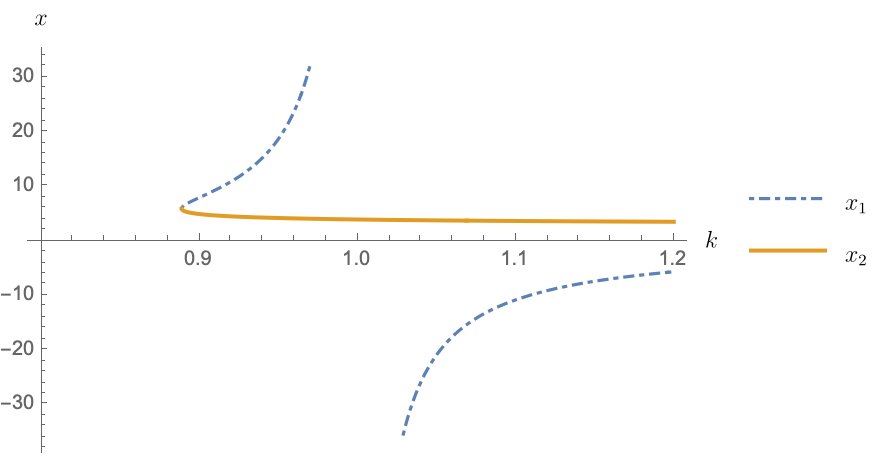}
	\caption{Nontrivial roots of the Schwarzschild limit are plotted as a function of the energy of the particle, $k$. In the interval $\frac{8}{9} \le k <1$,  $x_1$ is positive and it is negative and unphysical for $1<k$; $x_2$ is positive for $\frac{8}{9}\le k$. Both roots are located outside the event horizon. At $k=\frac{8}{9}$, $x_1=x_2=6$, corresponding to the ISCO.  }
	\label{fig:schwarzschildroots}
\end{figure}

\begin{figure}
	\centering
	\includegraphics[width=1\linewidth]{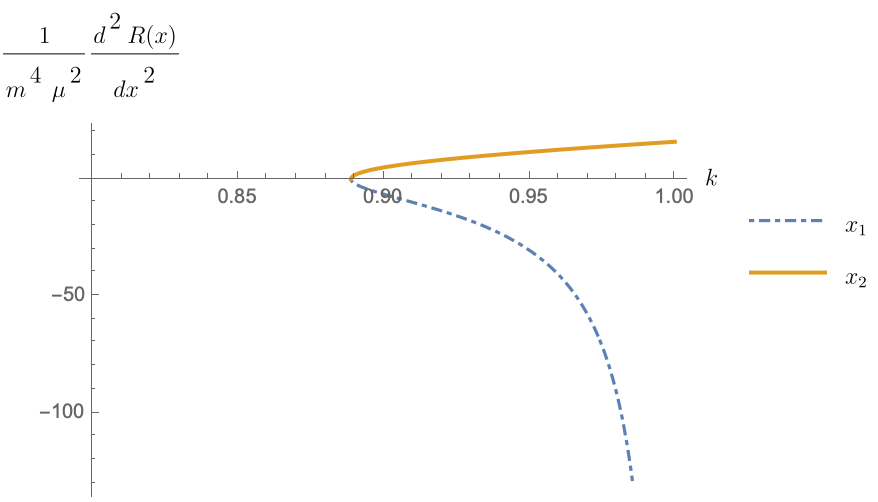}
	\caption{For the Schwarzschild black hole case, the second derivative of $R$ is plotted as a function of the energy of the particle, $k$. For the first root, the second derivative is negative, implying that this root corresponds to a stable orbit. For the second root, it is positive, implying that is an unstable orbit.}
	\label{fig:schwarzschildstability}
\end{figure}

\noindent Clearly, due to the spherical symmetry, there is no dependence on $\nu$.
$x_1(k)$ represents solutions that are positive only for a small energy range, $\frac{8}{9} \le k<1$ and negative and unphysical otherwise. In this range, $x_1(k)$ is outside the event horizon. On the other hand $x_2(k)$ represents solutions outside the event horizon for particles with energy $\frac{8}{9} \le k$.  For $k\rightarrow \infty$, $x_1$ goes to zero and $x_2 = 3$ which is the unstable photon ring.  In Fig. \ref{fig:schwarzschildroots}, we  plot the roots (\ref{Schw}) as a function of $k$ and in Fig. \ref{fig:schwarzschildstability}, 
we show the stability of $x_1$ and the  instability of $x_2$.  The roots coalesce  at $k= \frac{8}{9}$ for which $x_1= x_2 = 6$ corresponding to the inner most stable orbit (ISCO).

\section{Equatorial Orbits for the Kerr Black hole}

\begin{figure}[h]
	\centering
	\includegraphics[width=1 \linewidth]{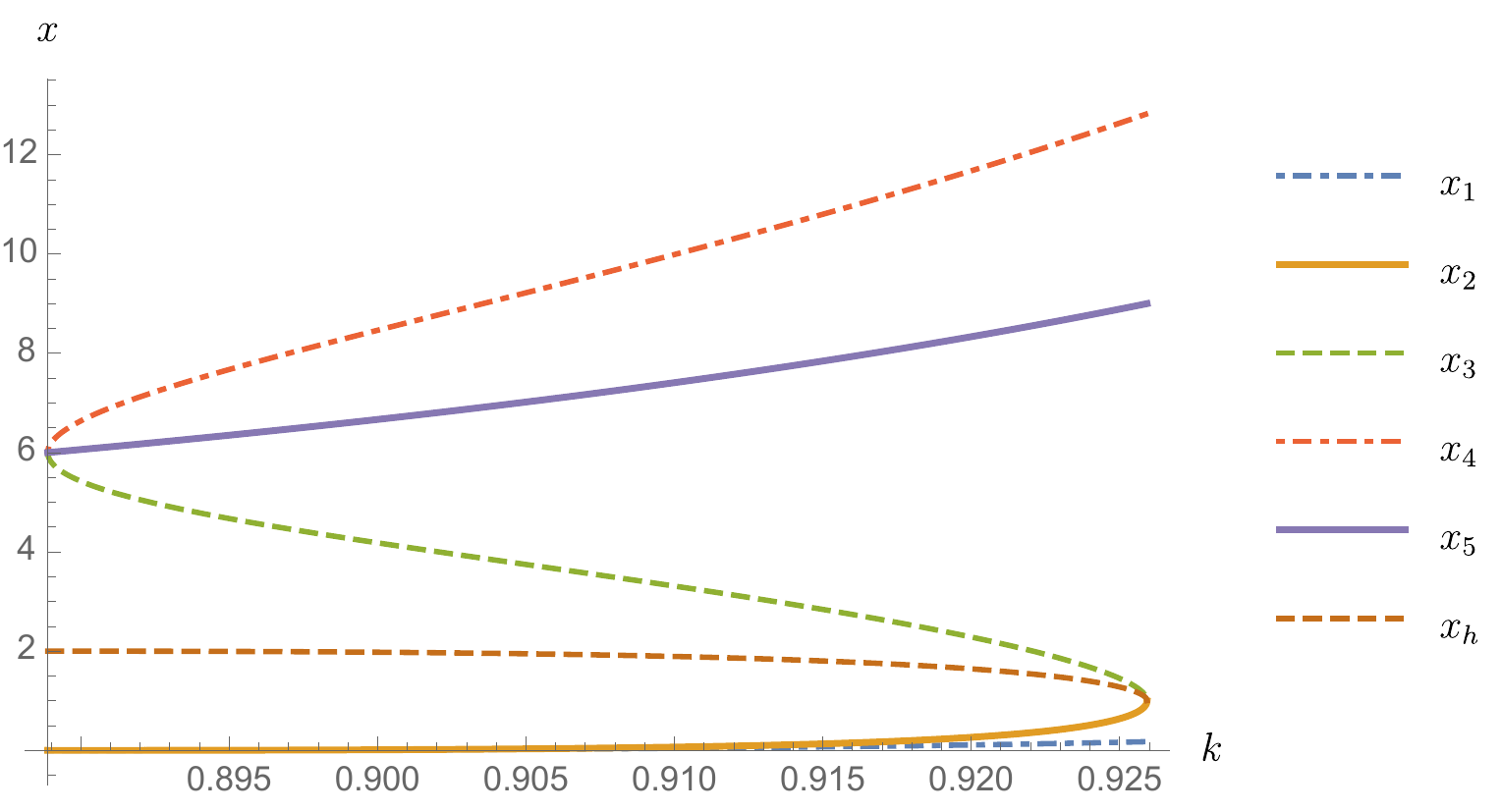}
	\caption{Real roots of the polynomial (\ref{polyeq}) and the event horizon are plotted as a function of the energy of the particle, $k$, at the equatorial plane of the Kerr black hole for the critical rotation parameter $u_{{\cal D}}$ given as (\ref{u_denk}).}
	\label{fig:roots-equator}
\end{figure}

\noindent Let us set $\nu =0$ in (\ref{polynomial}) which restricts the motion to the equatorial plane of the rotating black hole. Then (\ref{polynomial}) reduces to a sextic polynomial equation
\begin{eqnarray}\label{polyeq}
p(x,u,\nu=0,k)&=&(k-1)^2 x^6-2 (k-1) (3 k-4) x^5 \nonumber \\
&&+\left(9 k^2-32 k+24\right) x^4\\
&&-2 x^3 \left(2 k^2 u-3 k u-12 k+u+16\right) \nonumber \\
&&-2 x^2 (5 k u-4 u-8)-8 u x+u^2 =0, \nonumber 
\end{eqnarray}
which cannot be solved analytically in this generic form. Therefore, in order to simplify the calculations and make the polynomial factorizable, the values at which the discriminant of this polynomial with respect to $x$ 
\begin{eqnarray}\label{discriminant}
{\cal D}&=&4096 (k-1)^5 k^3 u^5 \left((k-1)^2 u+4 k\right)^2 \times\\
&&\left[729 (k-1)^4 u^3-27 (k (27 k-110)+91) (k-1)^2 u^2\right.\nonumber \\
&&\left.-8 (9 k (21 (k-4) k+101)-344) u-16 (8-9 k)^2\right] \nonumber,
\end{eqnarray}
vanishes can be found. First let us solve the discriminant equation as a function of $u=u(k)$, which yields a single  viable $u$
\begin{equation}
u=u_{{\cal D}}=\frac{4 (8-7 k)}{27 (1-k)^2}-\frac{16 \sqrt{2} \sqrt{\frac{k}{1-k}}}{27 (1-k)}.
\label{u_denk}
\end{equation}
For $u_{{\cal D}}$, the polynomial (\ref{polyeq}) can be factored as
\begin{equation}
p(x,u_{\cal_{D}},\nu=0,k)=-\frac{(3 k x-3 x+2)^2}{729 (k-1)^4}\times p_1(x,k),
\end{equation}
where $p_1(x,k)$ is a quartic polynomial which is somewhat cumbersome, but if one defines $k:= 1-\xi$, then it is simplified to the form
\begin{eqnarray}
p_1(x,\xi)&=&-81 \xi ^6 x^4+54\xi ^5\left(9\xi+1\right) x^3\\
&&-27\xi ^4\left(27 \xi ^2+18 \xi-1\right)x^2\nonumber\\
&&+\left[12 \xi ^3\left(654 \xi ^2-28 \xi-1\right)\right.\nonumber\\
&&\left.-96 \sqrt{2}(1-\xi)^{1/2}\xi ^{7/2} \left(1 -2\xi\right)\right] x\nonumber\\
&&-4\xi^2\left(17 \xi^2+46 \xi+1\right)\nonumber\\
&&-32 \sqrt{2} (1-\xi)^{1/2} \xi ^{5/2}\left(7\xi +1\right)\nonumber
\end{eqnarray}
 which has two real roots in the interval $\frac{1}{3}<k<\frac{8}{9}$ and four real roots in the interval $\frac{8}{9}<k<\frac{25}{27}$. Two of the roots of this quartic polynomial are always located inside the event horizon. The root of the quadratic part is real in the interval $\frac{1}{3}<k<\frac{25}{27}$. All the roots of the polynomial and the event horizon can be seen in Fig. \ref{fig:roots-equator}. 

\begin{figure}
	\centering
	\includegraphics[width=1\linewidth]{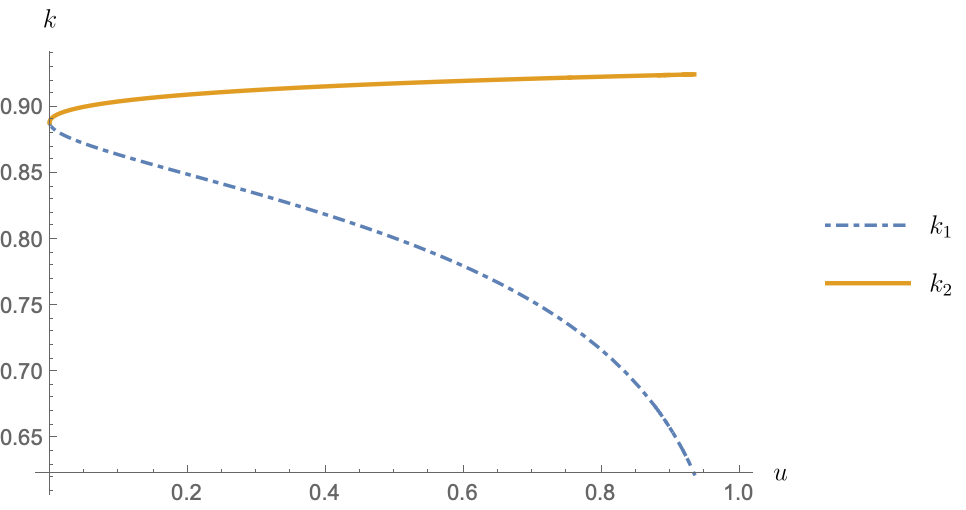}
	\caption{Two physically viable solutions, $k_1$ and $k_2$, at which the discriminant of the polynomial (\ref{discriminant}) vanishes are plotted as a function of the rotation parameter, $u$.}
	\label{fig:k6andk8}
\end{figure}

\noindent In addition to the above analysis where we took $k$ to be the independent variable,  to conform with our earlier work \cite{Aydin}, we would like to invert (\ref{u_denk}) and 
study the behavior of the roots as a function of the  black hole rotation parameter $u$. Even though, (\ref{u_denk}) is a quartic polynomial in $k$, there are only two physically viable solutions. Let us call them $k_1$ and $k_2$ which are found analytically, but we do not depict them here since they are rather long. All the ensuing discussion uses  these analytical expressions. Fig. \ref{fig:k6andk8} shows these two solutions as function of $u$.

\begin{figure}
	\centering
	\includegraphics[width=1\linewidth]{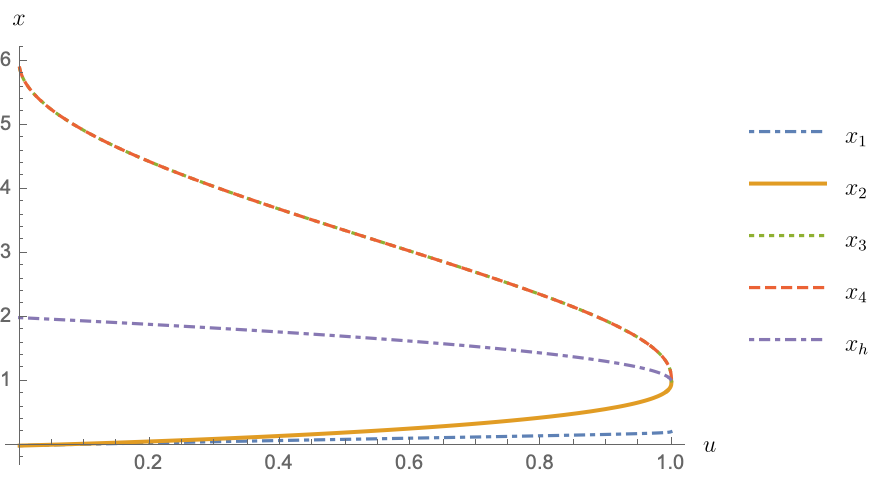}
	\caption{Real roots of the first solution ($k_1$) on the equatorial plane  are plotted as a function of the rotation parameter, $u$; roots $x_3$ and $x_4$, which coincide, are located outside the event horizon which is plotted as $x_h$.}
	\label{fig:rootsk6}
\end{figure}

\begin{figure}
	\centering
	\includegraphics[width=1\linewidth]{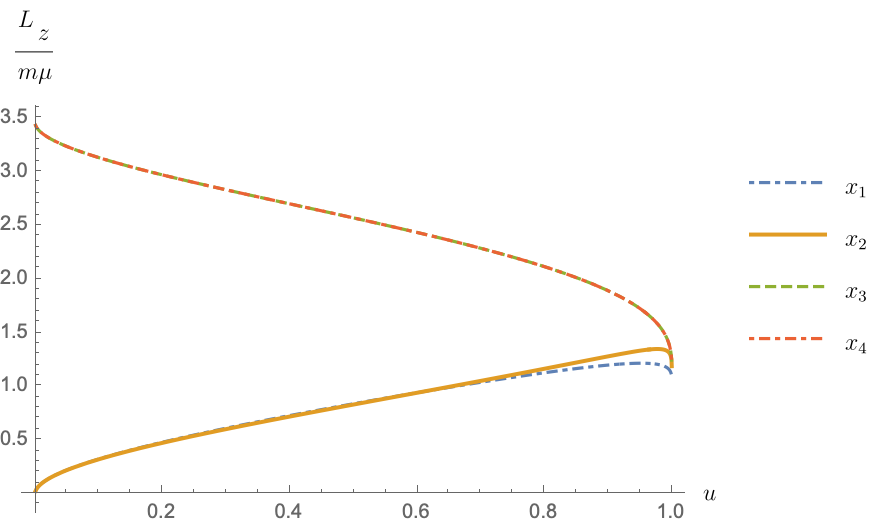}
	\caption{The $z$-components of the angular momentum per particle's mass and black hole mass of the roots of the first solution ($k_1$) on the equatorial plane are plotted as a function of the rotation parameter, $u$.  All the roots represent prograde orbits. The outer coinciding radii are monotonic in $u$, while the inner radii are nonmonotonic. }
	\label{fig:impactparameterk6}
\end{figure}

\begin{figure}
	\centering
	\includegraphics[width=1\linewidth]{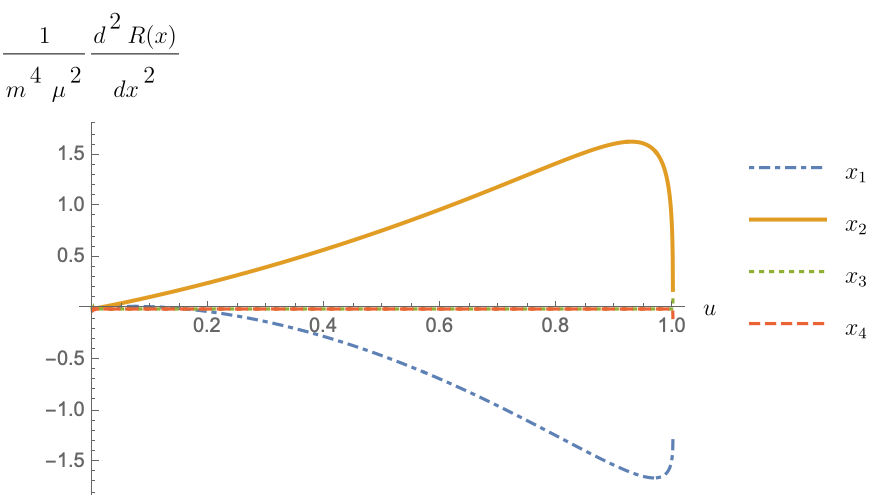}
	\caption{The second derivative of $R$ corresponding to the roots of the first solution ($k_1$) on the equatorial plane is plotted as a function of the rotation parameter, $u$. For inner roots, $x_1$ is a stable orbit and $x_2$ is an unstable orbit. The outer roots are saddle points.}
	\label{fig:secondderivativek6}
\end{figure}

\noindent For $k_1$, $p=p(x,u,\nu=0,k_1)$, the polynomial is factored and there will be six nontrivial solutions, four of which are real. The roots, together with the event horizon, are plotted as a function of the rotation parameter in Fig. \ref{fig:rootsk6}. $x_3$ and $x_4$ coincide and they are located outside of the event horizon but $x_1$ and $x_2$ are inside the event horizon. Fig. \ref{fig:impactparameterk6} shows that all of the roots represent prograde orbits. From Fig. \ref{fig:secondderivativek6}, one can see that the inner root $x_1$ represents a stable orbit while $x_2$ represents an unstable orbit. The outer roots are saddle points.

\begin{figure}
	\centering
	\includegraphics[width=1\linewidth]{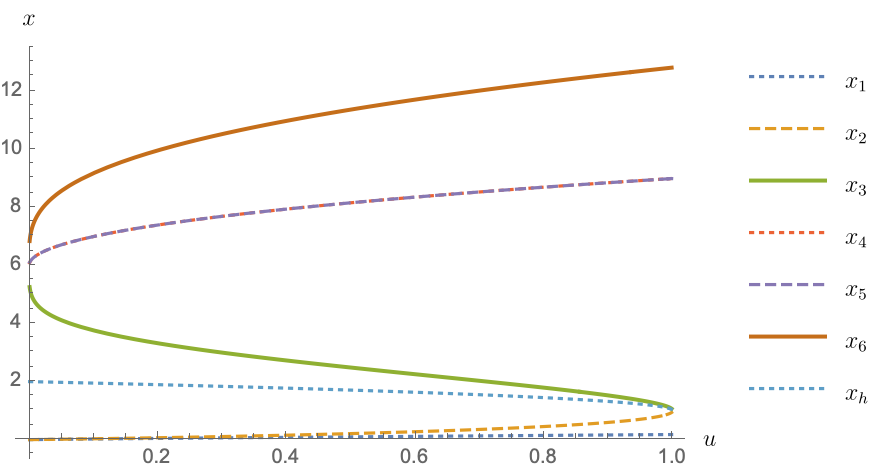}
	\caption{The real roots of the second solution ($k_2$) on the equatorial plane are plotted as a function of the rotation parameter, $u$. $x_1$ and $x_2$ are located inside the event horizon. $x_4$ and $x_5$ coincide.}
	\label{fig:rootsk8}
\end{figure}

\begin{figure}
	\centering
	\includegraphics[width=1\linewidth]{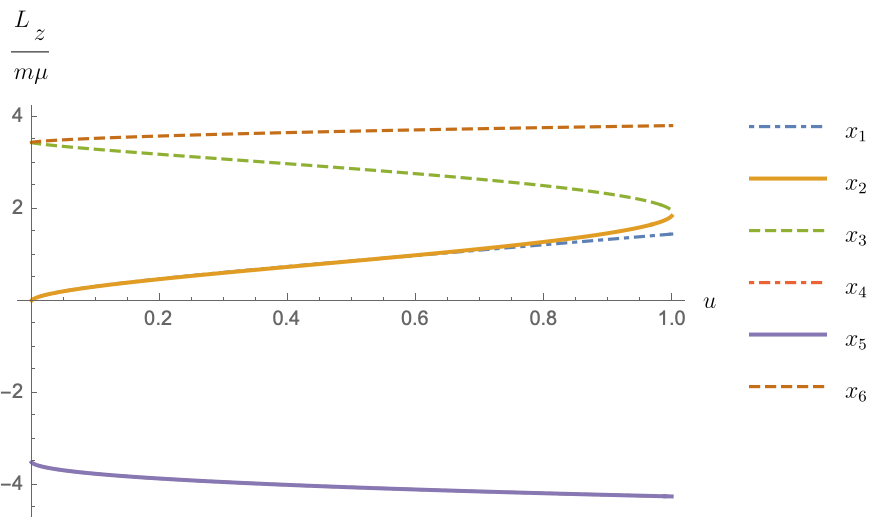}
	\caption{The $z$-components of the angular momentum per particle's mass and black hole mass of the roots of the second solution ($k_2$) on the equatorial plane are plotted as a function of the rotation parameter, $u$. It can be seen from this graph that roots $x_3$ and $x_6$ represent prograde orbits and $x_4$ and $x_5$ represent retrograde orbits.}
	\label{fig:impactparameterk8}
\end{figure}

\begin{figure}
	\centering
	\includegraphics[width=1\linewidth]{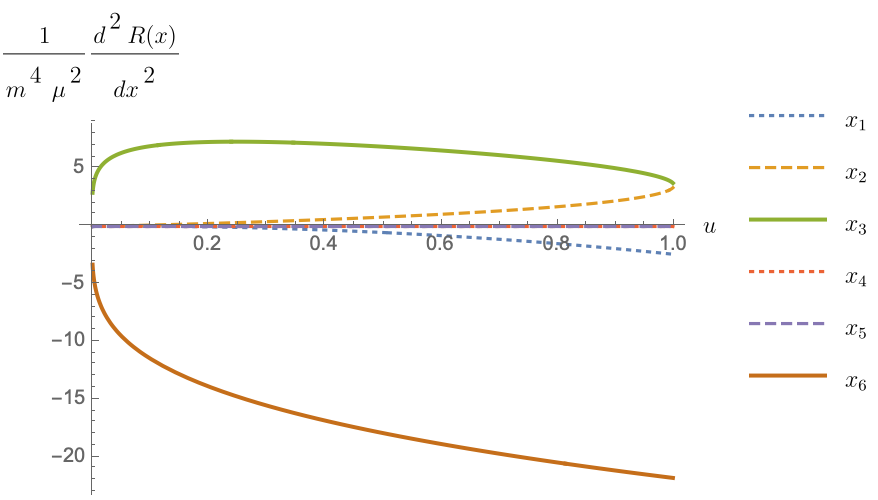}
	\caption{The second derivative of $R$ corresponding to roots of the second solution ($k_2$) on the equatorial plane is plotted as a function of the rotation parameter, $u$. $x_2$ and $x_3$ correspond to unstable orbits while $x_1$ and $x_6$ correspond to stable orbits. $x_4$ and $x_5$ are saddle points.}
	\label{fig:secondderivativek8}
\end{figure}

\noindent For $k_2$, $p=p(x,u,\nu=0,k_2)$, there are six nontrivial solutions all of which are real. The roots are plotted as a function of the rotation parameter as in Fig. \ref{fig:rootsk8}. $x_4$ and $x_5$ coincide; and $x_1$ and $x_2$ are inner roots. In Fig. \ref{fig:impactparameterk8}, one can see that $x_4$ and $x_5$ represent retrograde orbits while the rest of the roots represent prograde orbits. Fig. \ref{fig:secondderivativek8} shows that $x_1$ and $x_6$ correspond to stable orbits while $x_2$ and $x_3$ correspond to unstable orbits. The outer roots are saddle points.

\section{Zero Energy Solutions}

\begin{figure}[h]
	\centering
	\includegraphics[width=1 \linewidth]{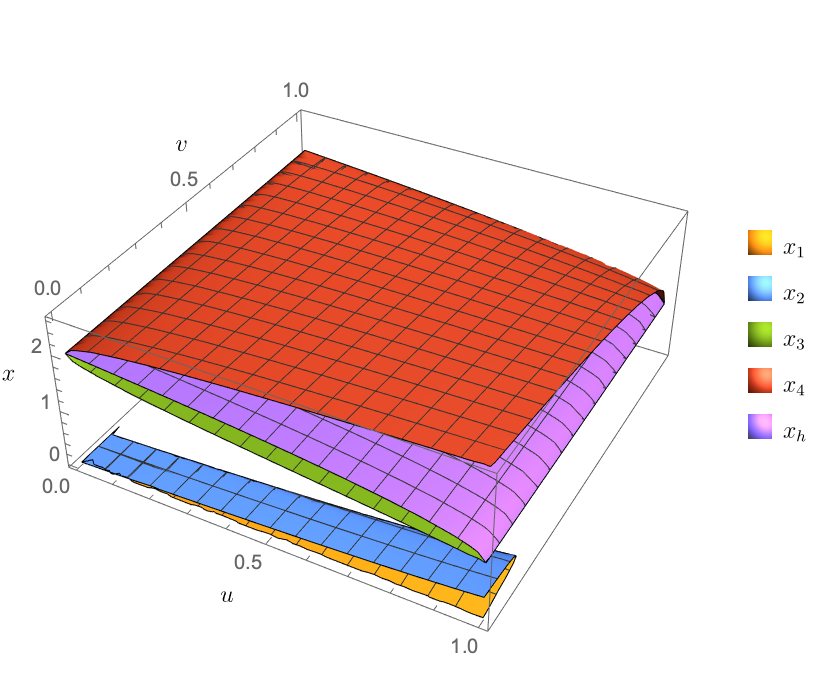}
	\caption{Behavior of the orbit of a particle with zero energy outside the event horizon as a  function of the rotation parameter, $u$, and  the inclination angle, $\nu$.}
	\label{fig:zero-energy-r1}
\end{figure}

\begin{figure}[h]
	\centering
	\includegraphics[width=1\linewidth]{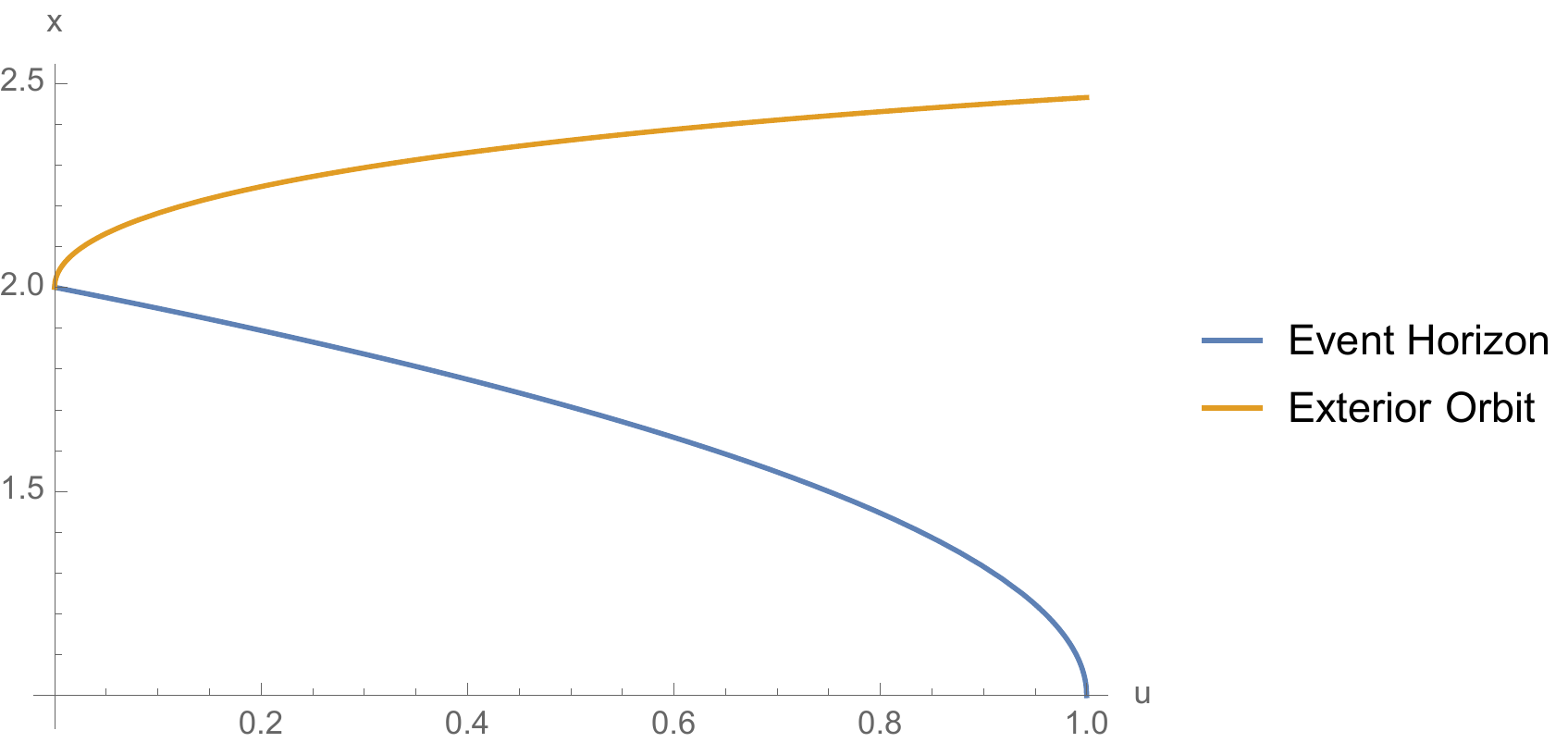}
	\caption{The exterior orbit and the event horizon are plotted as a function of rotation parameter, $u$, at $\nu=0.2$ plane.}
	\label{fig:zeroenergynu02}
\end{figure}

\begin{figure}[h]
	\centering
	\includegraphics[width=1\linewidth]{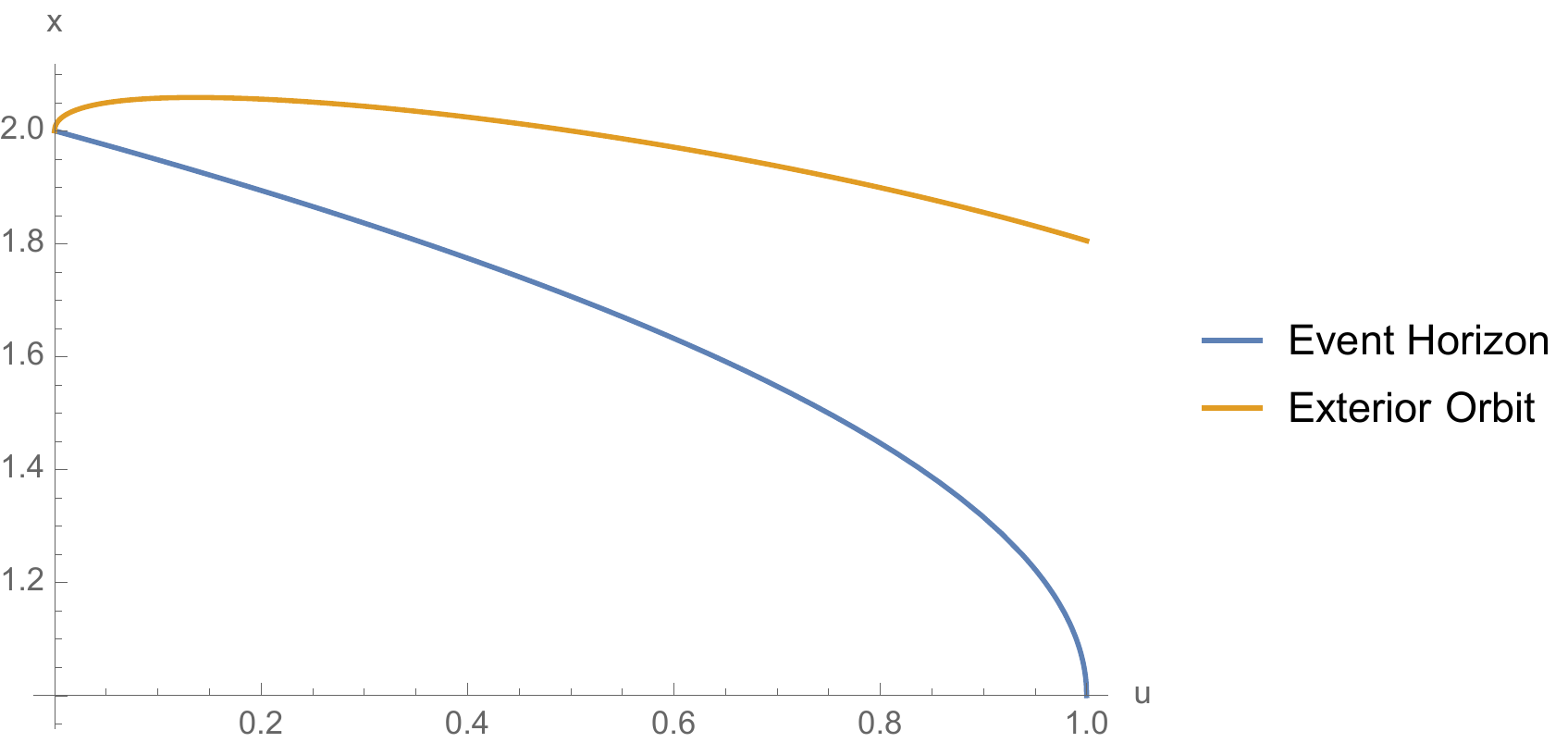}
	\caption{The exterior orbit and the event horizon are plotted as a function of the rotation parameter, $u$, at $\nu=0.8$. The dependence is nonmonotonic.}
	\label{fig:zeroenergynu08}
\end{figure}

\noindent Let us consider the zero energy case, $k=0$ for which  (\ref{polynomial}) simplifies to
\begin{equation}\label{zeropoly}
p(x)=x^2 \left(\nu  u^2+2(2+\nu  u) x^2-(3 \nu +1) u x+x^4-4 x^3\right)^2.
\end{equation}
This equation implies that there are four nontrivial orbits, one of which is located outside the event horizon as shown in Fig. \ref{fig:zero-energy-r1}.  One can find all the roots analytically but their depiction here is not needed since the expressions are rather complicated. Instead, we use these expressions to plot the figures. In Fig. \ref{fig:zeroenergynu02} and \ref{fig:zeroenergynu08}, the exterior root and the event horizon are plotted as functions of the rotation parameter, $u$, at different inclination angle values. It is important to state that at small inclination angles, such as $\nu=0.2$ as shown in Fig. \ref{fig:zeroenergynu02}, the exterior root has a monotonic behavior, but at large  inclination angles such as $\nu=0.8$, as shown in Fig. \ref{fig:zeroenergynu08}, the exterior root has a nonmonotonic behavior in $u$.

\noindent In order to understand this nonmonotonic behavior, the discriminant of the nontrivial part of the polynomial with respect to the rotation parameter can be investigated. The resulting equation becomes
\begin{equation}
{\cal D} = 4 \nu ^2 x^4-12 \nu ^2 x^3+9 \nu ^2 x^2-4 \nu  x^4+12 \nu  x^3-10 \nu  x^2+x^2.
\end{equation}
 Specific orbits as a function of $\nu$ such that $x=x(\nu)$, which make the discriminant, ${\cal D}$, vanish, can be calculated. There are two such roots:
\begin{eqnarray}
x_\pm(\nu) &=& \frac{3}{2}\pm\frac{1}{\sqrt{\nu}}, \,\,\,\ \nu \ne 0.
\end{eqnarray}
$x_+(\nu)$ lies outside the event horizon for the physically viable range ($0< \nu \le 1$) of the inclination angle. $x_+(\nu)$, evaluated at $\nu=0.8$,  gives the maximum value  $x_{\text{max}}=2.05902$; in other words, this is the farthest point to which the exterior orbit of the zero energy particle can reach. In order to find the corresponding rotation parameter, the polynomial equation (\ref{zeropoly}) should be recalculated with $x=x_+(\nu)$. The specific $u$ value which makes the resulting polynomial equation vanish is
\begin{equation}
u_+=u_+(\nu)=\frac{\nu \left(2-3 \sqrt{\nu}\right)+\sqrt{\nu}}{4 \nu^2}.
\end{equation}
$u_+(\nu)$ provides the turning point of $x$ and it can be called the critical rotation parameter. When it is evaluated at $\nu=0.8$, the result becomes $u=0.13586$. This result can also be observed in Fig. \ref{fig:zeroenergynu08}. When $u_+(\nu)$ is evaluated at $\nu=0.2$, the result becomes $u=3.61803$. This result is out of the physically meaningful range and that is why a nonmonotonicty is not observed in Fig. \ref{fig:zeroenergynu02}.

\section{Nonmonotonic Behavior at Nonzero Energy Values}

\noindent The nonmonotonic behavior observed for the zero energy particle solution at large inclination angle values in the previous section, can also be observed for particles with positive energy. Two exemplary cases are shown in the Table (\ref{table:nonmonotic}). In order to detect the turning point at which the nonmonotonic behavior starts, a similar procedure applied for the zero energy particle in the previous section can be followed. First, the discriminant of the polynomial equation (\ref{polynomial}) with respect to the rotation parameter, $u$, is calculated:
\begin{eqnarray}
\cal_{D}&=&256 k^2 (\nu -1)^2 \nu ^2 x^{18} (k x-x+1)^2\\
&&\times (3 k \nu  x+k x+4 \nu -3 \nu  x-x)^2\nonumber \\ 
&&\times \left(-4 k \nu +k \nu  x^2-k x^2-\nu  x^2+x^2+2 \nu  x-2 x\right)^2 \nonumber\\
&&\times \left( 256 k^2 \nu ^2+16 k^4 \nu ^2 x^6-64 k^3 \nu ^2 x^6\right.\nonumber\\
&&\left. \hskip 0.6 cm +96 k^2 \nu ^2 x^6+96 k^3 \nu ^2 x^5\right. \nonumber \\ 
&&\left. \hskip 0.6 cm -288 k^2 \nu ^2 x^5-96 k^4 \nu ^2 x^4+288 k^3 \nu ^2 x^4 \right.\nonumber\\
&&\left. \hskip 0.6 cm -72 k^2 \nu ^2 x^4-416 k^3 \nu ^2 x^3\right.\nonumber\\
&&\left. \hskip 0.6 cm+832 k^2 \nu ^2 x^3+144 k^4 \nu ^2 x^2-288 k^3 \nu^2 x^2\right.\nonumber\\
&&\left. \hskip 0.6 cm -456 k^2 \nu ^2 x^2-64 k^4 \nu  x^4\right.\nonumber\\
&&\left. \hskip 0.6 cm +192 k^3 \nu  x^4-200 k^2 \nu  x^4-192 k^3 \nu  x^3\right.\nonumber\\
&&\left. \hskip 0.6 cm +384 k^2 \nu  x^3-168 k^2 \nu  x^2\right. \nonumber \\
&&\left. \hskip 0.6 cm +384 k^3 \nu ^2 x-384 k^2 \nu ^2 x-64 k \nu ^2 x^6\right.\nonumber\\
&&\left. \hskip 0.6 cm +288 k \nu ^2 x^5-336 k \nu ^2 x^4 \right. \nonumber \\
&& \left. \hskip 0.6 cm -200 k \nu ^2 x^3+600 k \nu ^2 x^2+80 k \nu  x^4\right.\nonumber\\
&&\left. \hskip 0.6 cm -216 k \nu  x^3+168 k \nu  x^2\right.\nonumber\\
&&\left.\hskip 0.6 cm -288 k \nu ^2 x-32 k \nu  x +16 \nu ^2 x^6-96 \nu ^2 x^5\right.\nonumber\\
&&\left. \hskip 0.6 cm +216 \nu ^2 x^4-216 \nu ^2 x^3\right.\nonumber\\
&&\left.\hskip 0.6 cm+81 \nu ^2 x^2-8 \nu  x^4+24 \nu  x^3-18 \nu  x^2+x^2 \right) \nonumber.
\end{eqnarray}
Then, nontrivial $\nu=\nu(x,k)$ values which make the discriminant, $\cal_{D}$, vanish are calculated. The physically meaningful result is
\begin{equation}\label{critical}
\nu_{\text{critical}}=\frac{x^2}{A+B},
\end{equation}
where
\begin{eqnarray}
A &\equiv& x \left(4 \left((k-1)^2 (8 (k-1) k+1) x^3\right.\right.\\
&&\left.\left.\hskip 0.4 cm +3 (k-1) (8 (k-1) k+1) x^2\right.\right.\nonumber\\
&&\left.\left.\hskip 0.4 cm +21 (k-1) k x+4 k\right)+9 x\right), \nonumber
\end{eqnarray}
and
\begin{equation}
B \equiv -8 \sqrt{(1-2 k)^2 k (2 (k-1) x+3)^2 \left((k-1) x^2+x\right)^3}.
\end{equation}
At this moment, the turning point can be decided with the help of the $\nu_{\text{critical}}$ and the polynomial equation (\ref{polynomial}). For unit energy particles, $k=1$, the critical inclination angle becomes
\begin{equation}
\nu_{\text{critical}}=\frac{x^2}{x (9 x+16)-24 \sqrt{x^3}}.
\end{equation}
The $x$ values which make $\nu_{\text{critical}}=0.9$, as in the first case shown in the table, are $x=0.973493$ and $x=4.22548$. The second $x$ value represents a point outside the event horizon and corresponds to the point at which the maximum of the retrograde orbit occurs.  In other words, this is the farthest point a retrograde orbit for unit energy particles at the inclination angle $\nu=0.9$ can reach. The corresponding rotation parameter value can be calculated with the help of the polynomial equation (\ref{polynomial}) and it is $u=0.509021$. The result is consistent with Table (\ref{table:nonmonotic}).

\noindent Three important observations can be made by investigating the table. Firstly, this nonmonotonic behavior can only be observed at large inclination values; at small inclination angle, the nonmonotonicity occurs in the unphysical range of the rotation parameter, $u>1$. Secondly, the critical inclination angle decreases for higher energy values because as can be observed in (\ref{critical}), the critical inclination angle, $\nu_{\text{critical}}$, is inversely proportional to the energy parameter, $k$. Thirdly, the rotation parameter, $u$, corresponding to the turning point decreases for higher energy and higher inclination angles. In other words, The retrograde orbit starts to retreat at lower $u$ values for higher $k$ and $\nu$ values.
\begin{table}[h!]
	\caption{The prograde and retrograde orbits for different energy, $k$, inclination angle, $\nu$, and rotation parameter, $u$, values.}
	\label{table:nonmonotic}
	\centering
	{\begin{tabular}{||c c c c c||} 
			\hline
			$k$ & $\nu$ & $u$ & $x_{\text{prograde}}$ & $x_{\text{retrograde}}$ \\ 
			\hline
			1 & 0.9 & 0.1 & 3.74854 & 4.15469 \\ 
			\hline
			1 & 0.9 & 0.2 & 3.60934 & 4.19326 \\ 
			\hline
			1 & 0.9 & 0.3   & 3.48475 & 4.21283 \\ 
			\hline
			1 & 0.9 & 0.4  & 3.36502 & 4.22241 \\ 
			\hline
			1 & 0.9 & 0.5  & 3.2458 & 4.22546 \\ 
			\hline
			1 & 0.9 & 0.6  & 3.1241 & 4.22368 \\ 
			\hline
			1 & 0.9 & 0.7  & 2.99709 & 4.21809 \\ 
			\hline
			1 & 0.9 & 0.8  & 2.86137 & 4.2093 \\ 
			\hline
			1 & 0.9 & 0.9  & 2.71186 & 4.19774 \\ 
			\hline
			2 & 0.8 & 0.1  & 3.00717 & 3.38174 \\ 
			\hline
			2 & 0.8 & 0.2 & 2.88105 & 3.42079 \\ 
			\hline
			2 & 0.8 & 0.3 & 2.76773 & 3.44253 \\
			\hline
			2 & 0.8 & 0.4 & 2.65803 & 3.45524 \\
			\hline
			2 & 0.8 & 0.5 & 2.54757 & 3.46209 \\
			\hline
			2 & 0.8 & 0.6 & 2.43293 & 3.46466 \\
			\hline
			2 & 0.8 & 0.7 & 2.31031 & 3.46387 \\
			\hline
			2 & 0.8 & 0.8 & 2.17403 & 3.46032 \\
			\hline
			2 & 0.8 & 0.9 & 2.01281 & 3.45437 \\
			\hline
	\end{tabular}}
\end{table}

\section{The Unit Energy Case}

\noindent Another case which is important to investigate is the unit energy orbits: $k=1$,  for which (\ref{polynomial}) turns into
\begin{eqnarray}\label{polynomial2}
p(x)&=&x^8-8 x^7+ x^6 (2 (3 \nu -1) u+16)\\
&&-x^5 (8 \nu  u+8 u)\nonumber \\ 
&&+x^4 \left(9 \nu ^2 u^2-4 \nu  u^2+u^2\right)+x^3 \left(8 \nu  u^2-16 \nu ^2 u^2\right)\nonumber\\
&&+x^2 \left(6 \nu ^2 u^3-2 \nu  u^3\right)+\nu ^2 u^4\nonumber. 
\end{eqnarray}
Once again the generic analytical solutions are not available, so let us consider some specific cases.  
\subsection{Equatorial Plane Solutions}
\begin{figure}[h]
	\centering
	\includegraphics[width=1\linewidth]{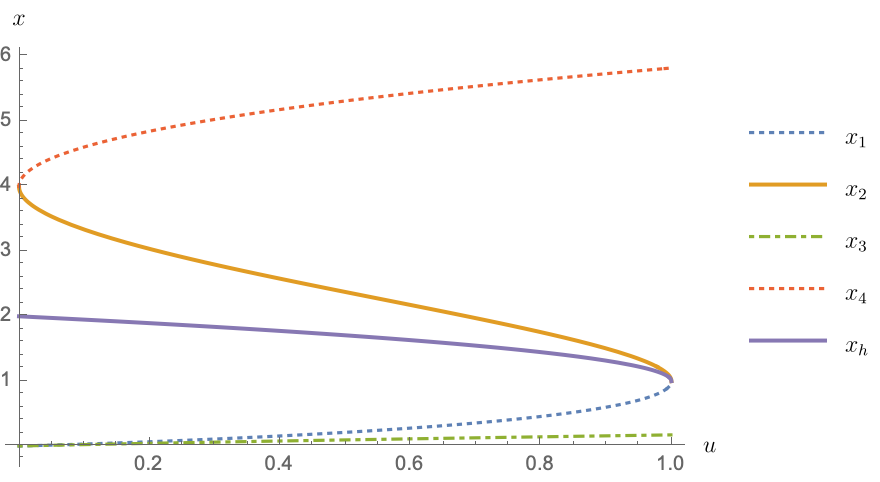}
	\caption{Roots of the polynomial on the equatorial plane for unit energy are plotted as a function of the rotation parameter.}
	\label{fig:unitenergyequatorialplaneroots}
\end{figure}

\noindent On the equatorial plane
\begin{equation}
\nu=0,
\end{equation}
at which the polynomial can be written in the form
\begin{eqnarray}
p(x)=p_1(x) \times p_2(x)
\end{eqnarray}
where
\begin{equation}
p_1(x)=x^4,
\end{equation}
and
\begin{equation}
p_2(x)=x^4-8 x^3+x^2\left(16-2 u\right)-8 u x+u^2
\end{equation}
of which the roots are
\begin{eqnarray}
&&x_1(u)=-\sqrt{u}-2 \sqrt{1-\sqrt{u}}+2,\nonumber \\
&&x_2(u)=-\sqrt{u}+2 \sqrt{1-\sqrt{u}}+2,\nonumber \\
&&x_3(u)=\sqrt{u}-2 \sqrt{\sqrt{u}+1}+2,\nonumber \\
&&x_4(u)=\sqrt{u}+2 \sqrt{\sqrt{u}+1}+2.
\end{eqnarray}
These are plotted as a function of the rotation parameter in Fig. \ref{fig:unitenergyequatorialplaneroots}.  The roots $x_2$ and $x_4$ are located outside of the event horizon. With the help of these solutions, the Lagrange-B\"{u}rmann method can be used to approximate solutions out of the equatorial plane for particles with unit energy. In Appendix B, we briefly describe this useful theorem. 

\subsection{Lagrange-B\"{u}rmann Method}

\noindent The polynomial found in (\ref{polynomial2}) is order of $2$ with respect to $\nu$. Hence, it has two solutions which are
\begin{eqnarray}
\nu_1&=&\frac{1}{u^2 \left(u^2+6 u x^2+(9 x-16) x^3\right)} \\
&&\times \left( u x^2 \left(u-x^2\right) (u+x (3 x-4))\right.\nonumber\\
&&\left.\hskip 0.6 cm -8 \sqrt{u^2 x^7 (u+(x-2) x)^2}\right) , \nonumber
\end{eqnarray}
and
\begin{eqnarray}
\nu_2&=&\frac{1}{u^2 \left(u^2+6 u x^2+(9 x-16) x^3\right)}\\
&&\times \left(8 \sqrt{u^2 x^7 (u+(x-2) x)^2}\right.\nonumber\\
&&\left.\hskip 0.6 cm +u x^2 \left(u-x^2\right) (u+x (3 x-4))\right).\nonumber
\end{eqnarray}
$\nu_2$ is the root which approaches to $0$ when $x=x_2$ or $x=x_4$. Therefore, it can be written that
\begin{equation}
\nu=\nu_2=f(x)
\end{equation}
and the Lagrange-B\"{u}rmann method allows us to estimate a function of the form
\begin{equation}
x=g(\nu)
\end{equation}
where
\begin{eqnarray}\label{LB}
g(\nu)&=&x_2+\left(\nu-f(x_2)\right)\lim_{x \to x_2}\left(\frac{x-x_2}{f(x)-f(x_2)}\right)\\
&&+\frac{\left(\nu-f(x_2)\right)^2}{2}\lim_{x \to x_2}\left(\frac{d}{dx}\left(\frac{x-x_2}{f(x)-f(x_2)}\right)^2\right). \nonumber
\end{eqnarray}
Two conditions of the Lagrange-B\"{u}rmann inversion theorem, namely $f(x_2)$ should be analytic and $f'(x_2)$ should be nonzero, are satisfied. The error rate of the approximate solutions for different values of rotation parameter at different inclination angles can be seen in the Table \ref{table:LBk1}. As can be observed, the approximation works quite well near the equatorial plane for the whole range of $u$. Nevertheless, near the polar plane, a better approximation is needed build on the $\nu=1$ solution.
\begin{table}[h!]
	\caption{The error rates of the approximate solution obtained with the Lagrange-B\"{u}rmann method to the unit energy particle orbits.}
	\label{table:LBk1}
	\centering
	{\begin{tabular}{||c c c||} 
			\hline
			$\nu$ & $u$ & $\text{Error (\%)}$ \\ 
			\hline
			0.1 & 0.1 & 0.00128664    \\ 
			\hline
			0.1 & 0.9 & 0.0298184   \\ 
			\hline
			0.9 & 0.1 & 2.28609    \\ 
			\hline
			0.9 & 0.9 & 20.3042    \\ 
			\hline
	\end{tabular}}
\end{table}

\section{Conclusions}

\noindent The method developed for spherical photon orbits in \cite{Aydin} has been extended to cover timelike orbits of massive particles around the Kerr black hole.  Orbits  on the equatorial plane have been investigated in detail. For nonequatorial and high inclination angles, a nonmonotonicity of the radius as a function of the black hole rotation parameter has been observed in the retrograde orbits for both zero and nonzero energy particles. The physical meaning of this is that around higher spinning black holes, the capture cross section of these particles is reduced for retrograde orbits. This fact is relevant for finding the maximum spin of a black hole with a thick accretion disk.  A formula has been derived to detect the critical inclination angle above which the nonmonotonicity sets in. We also used the Lagrange-B\"{u}rmann method to develop approximate analytical solutions around known exact solutions.

\section{Appendices}

\subsection{Revisiting the Photon Case}

\noindent For completeness, we would like to revisit some of the results of \cite{Aydin} [see also \cite{Aydin_Tez}]. For the spherical photon orbit at the equatorial plane, there are three solutions which are
\begin{eqnarray}
&& x_\pm = 4\cos^2 \left(\frac{1}{3} \cos ^{-1}(\pm \sqrt{u}  )\right), \\
&& x_{\text{in}}= 4 \sin ^2\left(\frac{1}{3} \sin ^{-1}\sqrt{u}\right).
\end{eqnarray}
Here, $x_+$ represents the retrograde orbit while $x_-$ represents the prograde orbit. $x_{\text{in}}$ is the orbit located inside the event horizon.

\noindent At the polar plane, there are only two physically viable solutions, which are
\begin{equation}
x_P= 1+2 \sqrt{1-\frac{u}{3}} \cos \left \{\frac{1}{3} \cos ^{-1}\left(\frac{1-u}{\left(1-\frac{u}{3}\right)^{3/2}}\right)\right \},
\end{equation}
and
\begin{equation}
x_{\text{in}}=1-2 \sqrt{1-\frac{u}{3}} \sin \left \{\frac{1}{3} \sin ^{-1}\left(\frac{1-u}{\left(1-\frac{u}{3}\right)^{3/2}}\right)\right \}.
\end{equation}

\begin{figure}[ht!]
	\centering
	\includegraphics[width=1\linewidth]{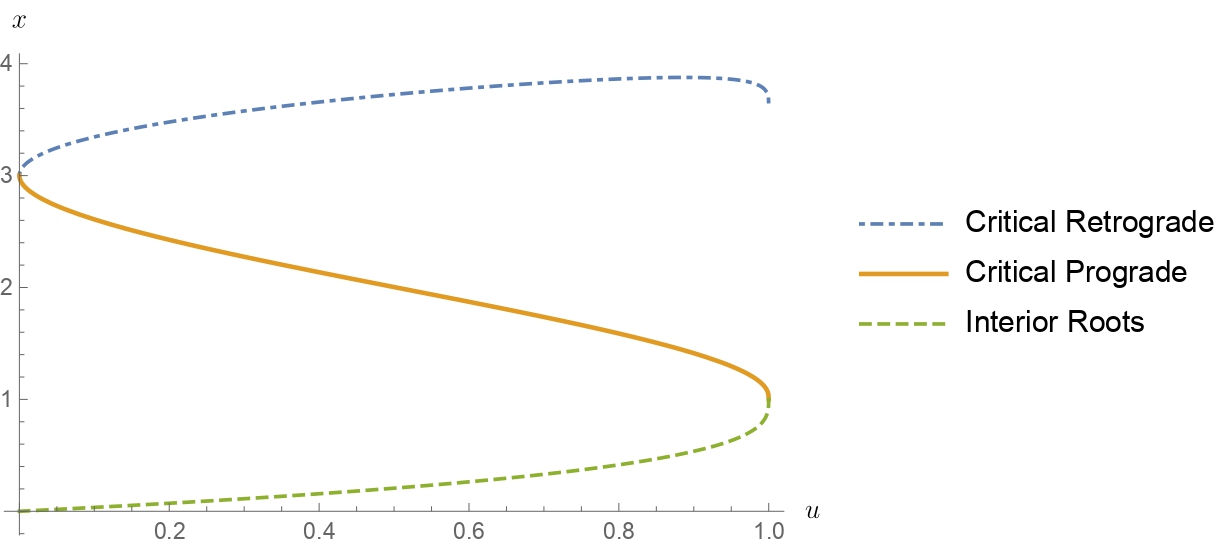}
	\caption{Real roots are plotted as a function of the rotation parameter at the critical inclination angle. }
	\label{fig:4criticalroots}
\end{figure}

\noindent In \cite{Aydin}, in addition to equatorial and polar solutions, a new exact solution which is located at the critical plane was found. At the critical plane given in (\ref{criticalphoton}), four real orbits can be found, two coincide, and lie inside the event horizon. The exact expressions are cumbersome but we can plot them as a function of the rotation parameter as in Fig. \ref{fig:4criticalroots}. The root  corresponding to the retrograde orbit  is nonmonotonic in $u$ while the other roots are monotonic.

\subsection{ Lagrange-B\"{u}rmann Theorem}

\noindent Lagrange-B\"{u}rmann theorem (or Lagrange-B\"{u}rmann inversion theorem) is a Taylor series expansion of the inverse of an analytical function. The theorem first published by Joseph-Louis Lagrange in 1770 \cite{Grossman}. Then, it was generalized by Hans Heinrich B\"{u}rmann in 1799 \cite{Whittaker}. 

\noindent Assume that there is a real analytic function, $f$, of the form 
\begin{equation}
\omega=f(z),
\end{equation}
where $\omega, z \in \mathbb{R}$. Assume also that there is a point $z_0$ at which the function is analytic,
\begin{equation}
\omega_0=f(z_0),
\end{equation}
and has a non-vanishing first derivative,
\begin{equation}
\frac{df}{dz}\Bigr|_{\substack{z=z_0}} \neq 0.
\end{equation}
Let $g$ be a function defined as the inverse of the function $f$,
\begin{equation}
z=g(\omega).
\end{equation}
Then, Lagrange-B\"{u}rmann theorem states that the inverse function can be expanded as
\begin{equation}
g(z)=\omega_0+\sum_{n=1}^{\infty} \omega_n \frac{\left(f(z)-f(z_0)\right)^n}{n!},
\end{equation}
where the coefficients are given by
\begin{equation}
\omega_n=\lim_{z\to z_0} \frac{d^{n-1}}{dz^{n-1}}\left(\frac{z-z_0}{f(z)-f(z_0)}\right)^n
\end{equation}
\cite{Whittaker}, \cite{Abramowitz}.

\subsection{Discriminant of higher order  polynomials}

\noindent For a polynomial of the form
\begin{equation}
p(x)=a_n x^n + a_{n-1} x^{n-1} + \dots + a_1 x + a_0,
\end{equation}
where $n>0$, the discriminant can be defined as
\begin{equation}
\mathcal{D}=a_n^{2n-2} \prod_{i, j}^{n} \left(r_i-r_j\right)^2,
\end{equation}
for $i<j$, where $r_i$'s are the roots of the polynomial. The discriminant vanishes only when two or more roots are equal, as can be observed from the definition. But the above definition requires the knowledge of all the roots, which is not generically available. Hence, we used the following  alternative definition. 

\noindent The discriminant of a polynomial of the form
\begin{equation}
p(x)=x^n + a_{n-1} x^{n-1} + \dots + a_1 x + a_0,
\end{equation}
where $n>0$, can also be defined as
\begin{equation}
\mathcal{D}=\left(-1\right)^{\frac{n\left(n-1\right)}{2}} \mathcal{R}\left(p(x),p'(x)\right)
\end{equation}
where $\mathcal{R}$ is the resultant \cite{Akritas}. 

\subsection{Generic Energy Values}

\noindent For energy values other than zero and unity, the polynomial (\ref{polynomial}) cannot be solved analytically in terms of a finite number of radicals even on the equatorial plane. Hence, in order to investigate the solutions with these energy values, a different approach should be devised, which we do below.

\subsubsection{Bound Orbits}

\noindent Numerical analysis can be used to observe the behavior of bound orbits with energy values between $k=0$ and $k=1$. Solutions on the equatorial plane are considered first. 

\begin{figure}[h]
	\centering
	\includegraphics[width=1\linewidth]{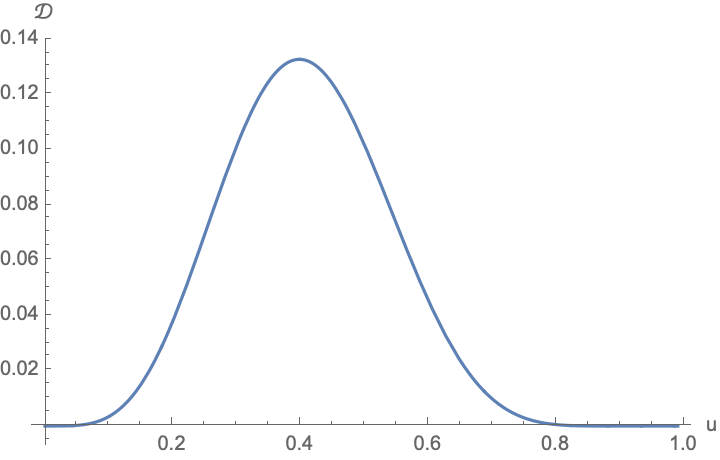}
	\caption{The graph of the discriminant of the polynomial with $\nu=0$ and $u=0.1$.}
	\label{fig:discriminant1}
\end{figure}

\noindent As a first case, the rotation parameter is set equal to $0.1$, $u=0.1$, and the energy value is gradually decreased starting with the unit energy, $k=1$. With the conditions $\nu=0$ and $u=0.1$, the polynomial equation (\ref{polynomial}) can be factored as
\begin{equation}
p(x)=x^4 \times p_1(x)
\end{equation}
where
\begin{eqnarray}
p_1(x)&=&x^6\left(k^2-2k+1\right)\\
&&+x^5\left(-6k^2+14k-8\right)\nonumber \\
&&+ x^4\left(9k^2-32k+24\right)\nonumber\\
&&+x^3\left(-0.4k^2+24.6k-32.2\right)\nonumber\\
&&+x^2 \left(16.8-k\right)-0.8x+0.01 \nonumber
\end{eqnarray}
The discriminant of $p_1(x)$ with respect to $x$ has roots at $k=0.864513$ and $k=0.904746$. There are no real roots of the polynomial located outside the event horizon for the interval where $0 < k < 0.864513$. There are two real roots in the interval $0.864512 < k < 0.904746 $, and four real roots in the interval $0.904746 < k < 1$ of the polynomial which represent orbits located outside the event horizon. The general behavior of the discriminant can be seen in Fig. \ref{fig:discriminant1}.

\begin{figure}[h]
	\centering
	\includegraphics[width=1\linewidth]{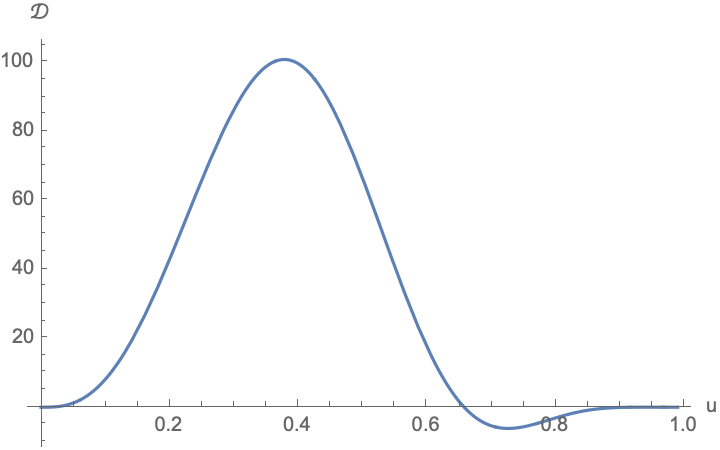}
	\caption{The graph of the discriminant of the polynomial with $\nu=0$ and $u=0.9$.}
	\label{fig:discriminant2}
\end{figure}

\begin{figure}[h]
	\centering
	\includegraphics[width=1\linewidth]{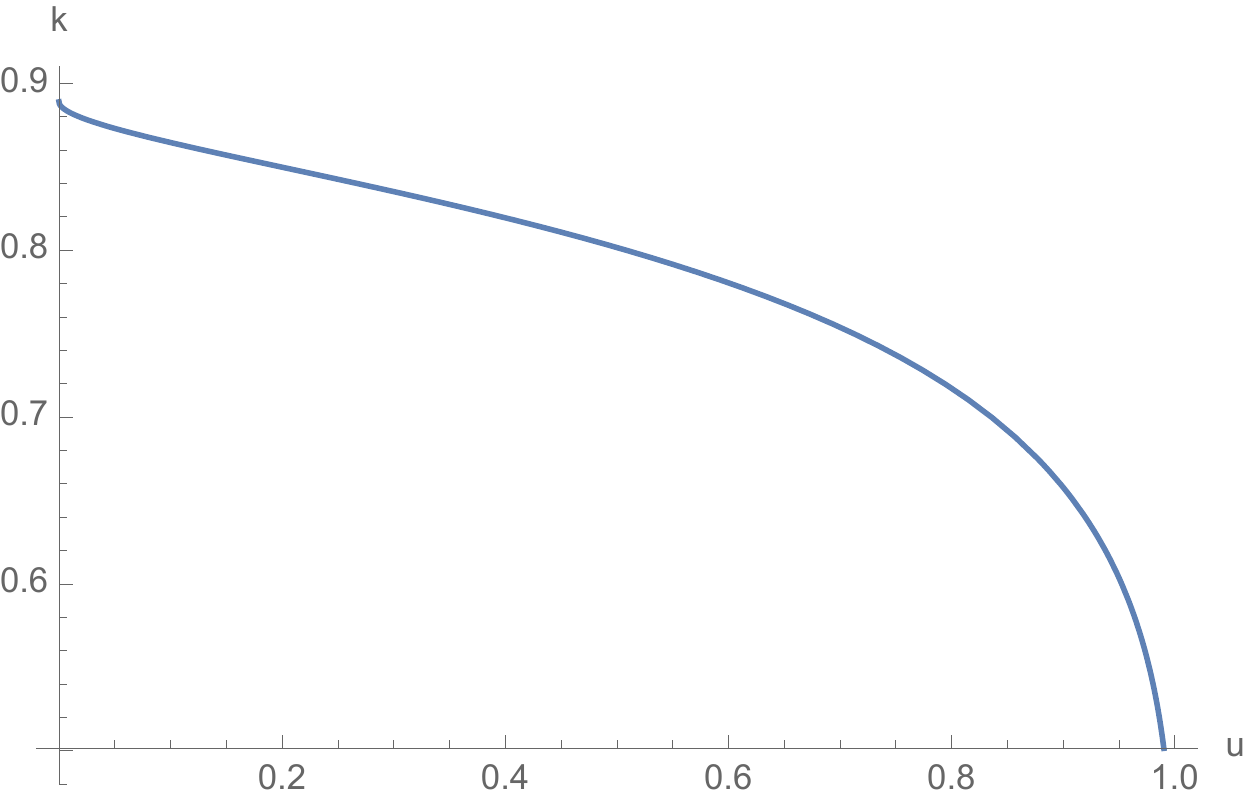}
	\caption{The graph of energy values, $k$, which make the discriminant vanish on the equatorial plane with respect to $u$.}
	\label{fig:kvsueq}
\end{figure}

\noindent As a second case, the rotation parameter is set equal to 0.9, $u=0.9$, and the energy value is gradually decreased by starting with the unit energy, $k=1$. This time, the discriminant of the polynomial equation vanishes at $k=0.657945$ and $k=0.924715$. There are no real roots located outside the event horizon for the interval $0 < k < 0.657945$. There are two real roots in the interval $0.657945 < k < 0.924715$, and there are four real roots in the interval $0.924715 < k < 1$ of the polynomial located outside the event horizon. The behavior of the discriminant can be seen in Fig. \ref{fig:discriminant2} . In Fig. \ref{fig:kvsueq}, the distribution of the energy values, $k$, which makes the discriminant vanish for all $u$ values on the equatorial plane can be seen.  

\noindent As an another method to investigate the behavior of equatorial orbits with energy values between $k=0$ and $k=1$, the Lagrange-B\"{u}rmann method can be applied to the exact solution obtained for the unit energy at the equatorial plane in order to approximate the solution in the desired intervals where there is a real root outside the event horizon. The error rates for some specific cases are summarized in Table \ref{table:LBk1eq}.
\begin{table}[h!]
	\caption{The error rates of the approximate solution obtained with the Lagrange-B\"{u}rmann method at the equatorial plane $\nu=0$ for bound orbits.}
	\label{table:LBk1eq}
	\centering
	{\begin{tabular}{||c c c||} 
			\hline
			$u$ & $k$ & $\text{Error (\%)}$ \\ 
			\hline
			0.1 & 0.9 & 3.03788    \\ 
			\hline
			0.9 & 0.9 & 0.0726006  \\ 
			\hline
			0.1 & 0.87 & 11.8191    \\ 
			\hline
			0.9 & 0.87 & 0.173522   \\ 
			\hline
	\end{tabular}}
\end{table}

\begin{figure}[h]
	\centering
	\includegraphics[width=1\linewidth]{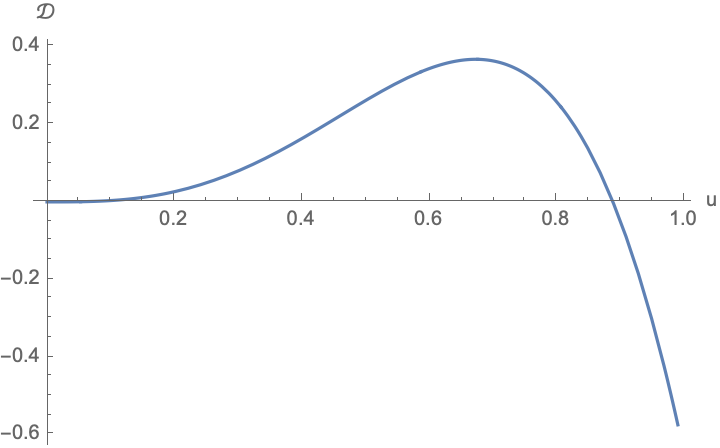}
	\caption{The graph of the discriminant of the polynomial with $\nu=1$ and $u=0.1$.}
	\label{fig:discriminant3}
\end{figure}

\noindent Now, the solutions on the polar plane can be investigated. With the condition $\nu=1$, the polynomial equation can be factored as 
\begin{equation}
p(x)=p_1^2(x)
\end{equation}
where
\begin{eqnarray}
p_1(x)&=&x^5(k-1)+x^4(4-3k)\\
&&+x^3 (2 k u-2 u-4)\nonumber\\
&&+x^2 (4 u-2 k u)+x \left(k u^2-u^2\right)+k u^2 \nonumber
\end{eqnarray}
As a first case, the rotation parameter is set equal to 0.1, $u=0.1$, and the energy value is gradually decreased by starting with the unit energy, $k=1$. The discriminant of $p_1(x)$ with respect to $x$ has a root at $k=0.888049$. There are real roots of the polynomial located outside of the event horizon for the interval where $0.888049<k \leq 1$ and for the special case  $k=0$. This time, there are other real roots of the discriminant in the interval $0 \leq k < 0.888049$, but they only affect the number of real orbits located inside the event horizon. The behavior of the discriminant can be seen in Fig. \ref{fig:discriminant3}.

\begin{figure}[h]
	\centering
	\includegraphics[width=1\linewidth]{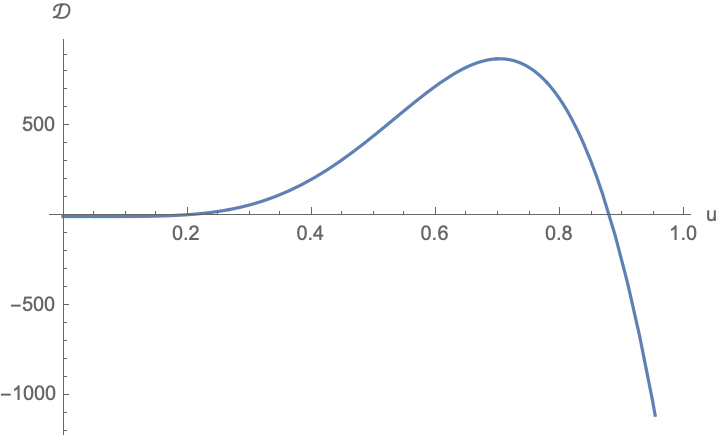}
	\caption{The graph of the discriminant of the polynomial with $\nu=1$ and $u=0.9$.}
	\label{fig:discriminant4}
\end{figure}

\begin{figure}[h]
	\centering
	\includegraphics[width=1\linewidth]{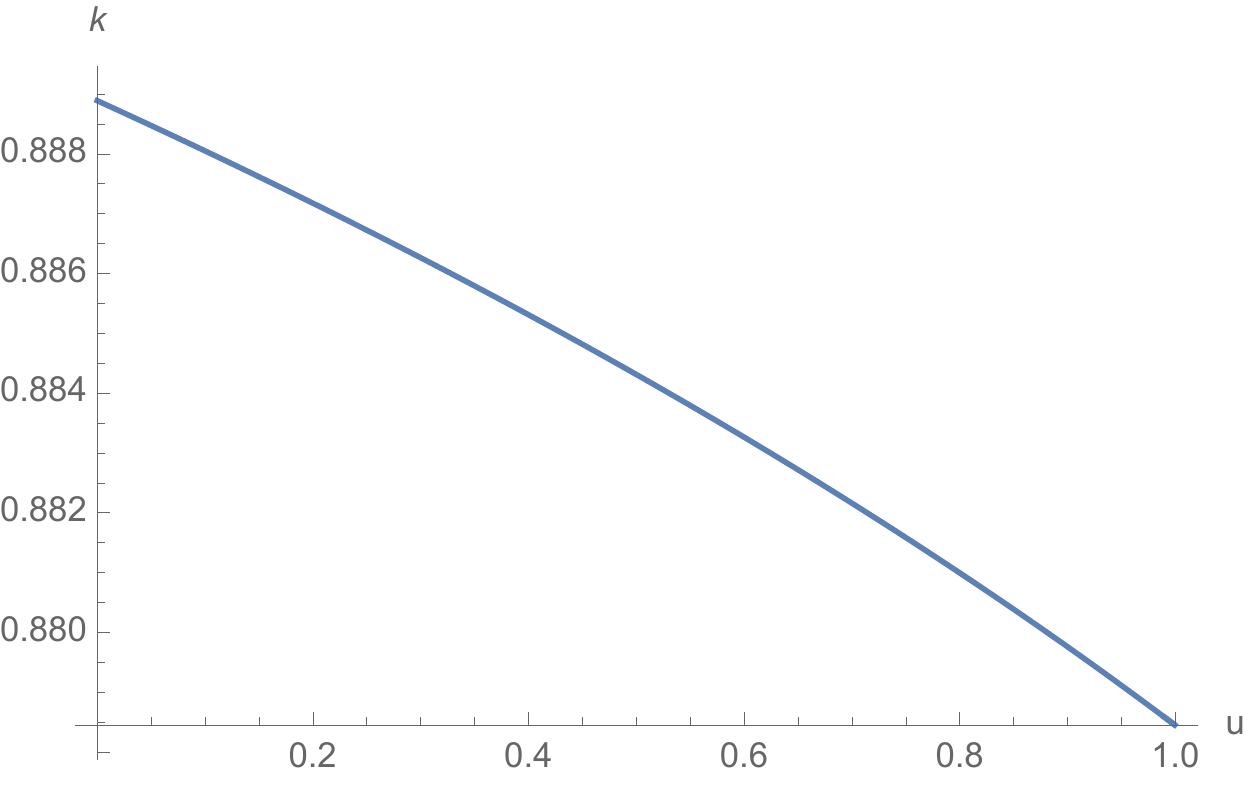}
	\caption{The graph of energy values, $k$, which makes the discriminant vanish on the polar plane with respect to $u$.}
	\label{fig:kvsupol}
\end{figure}

\noindent As a second case, the rotation parameter is set equal to 0.9, $u=0.9$ and the energy value is gradually decreased by starting with the unit energy, $k=1$. The discriminant of the polynomial equation vanishes at the point where $k=0.879759$. There are real roots located outside the event horizon for the interval $0.879759<k \leq 1$ and for the special case $k=0$. The general behavior of the discriminant can be observed in Fig. \ref{fig:discriminant4}.  In Fig. \ref{fig:kvsupol}, the distribution of the energy values, $k$, which make the discriminant vanish for all $u$ values on the polar plane can be seen.

\noindent Approximate solutions to investigate the behavior of polar orbits with energy values which are in the interval where there is a real root outside of the event horizon, can be obtained with the help of the Lagrange-B\"{u}rmann method. The error rates for some specific cases are shown in Table (\ref{table:LBk1pol}).
\begin{table}[h!]
	\caption{The error rates of the approximate solution obtained with the Lagrange-B\"{u}rmann method at the polar plane $\nu=1$ for bound orbits.}
	\label{table:LBk1pol}
	\centering
	{\begin{tabular}{||c c c||} 
			\hline
			$u$ & $k$ & $\text{Error (\%)}$ \\ 
			\hline
			0.1 & 0.9 & 2.9904    \\ 
			\hline
			0.9 & 0.9 & 1.70533  \\ 
			\hline
			0.1 & 0.89 & 9.12136    \\ 
			\hline
			0.9 & 0.89 & 4.01093   \\ 
			\hline
	\end{tabular}}
\end{table}

\subsubsection{Unbound Orbits}

\noindent The approach used for bound orbits with energy values $k<1$ can be used for unbound orbits with energy values $k>1$. As a first case, equatorial plane solutions are investigated. This time there is no limit observed on the energy levels. In other words, there always exist real orbits outside the event horizon for energy values $k>1$. The exact solutions obtained on the equatorial plane can be used to approximate solutions on planes with the critical inclination angle greater than zero, $\nu>0$ with the help of the Lagrange-B\"{u}rmann method. The error rates of some specific cases are summarized in Table (\ref{table:LBkGt1eq}).
\begin{table}[h!]
	\caption{The error rates of the approximate solution obtained with the Lagrange-B\"{u}rmann method at the equatorial plane, $\nu=0$ for unbound orbits.}
	\label{table:LBkGt1eq}
	\centering
	{\begin{tabular}{||c c c||} 
			\hline
			$u$ & $k$ & $\text{Error (\%)}$ \\ 
			\hline
			0.1 & 1.1 & 1.04468    \\ 
			\hline
			0.9 & 1.1 & 0.0477497  \\ 
			\hline
			0.1 & 1.5 & 24.1319    \\ 
			\hline
			0.9 & 2 & 19.2937   \\ 
			\hline
	\end{tabular}}
\end{table}

\noindent As a second case polar plane solutions are investigated. Again, there is no limit on $k$. The exact solutions of the polar plane for $k=1$ can be used to approximate solutions on planes with critical inclination angle $\nu<1$ with the help of the Lagrange-B\"{u}rmann method. The error rates for some specific cases are shown in Table (\ref{table:LBkGt1pol}).
\begin{table}[h!]
	\caption{The error rates of the approximate solution obtained with the Lagrange-B\"{u}rmann method at the polar plane, $\nu=1$ for unbound orbits.}
	\label{table:LBkGt1pol}
	\centering
	{\begin{tabular}{||c c c||} 
			\hline
			$u$ & $k$ & $\text{Error (\%)}$ \\ 
			\hline
			0.1 & 1.1 & 0.536127    \\ 
			\hline
			0.9 & 1.1 & 0.392406  \\ 
			\hline
			0.1 & 1.2 & 25.2596    \\ 
			\hline
			0.9 & 1.2 & 18.784   \\ 
			\hline
	\end{tabular}}
\end{table}
Higher error rates for some values do not deter us the least since we only took the first order term in the Lagrange-B\"{u}rmann expansion. One gets much better results by incorporating  more terms. But here we just wanted to demonstrate the  power of this technique.

\end{document}